\documentclass{sigchi}

\CopyrightYear{2017}
\setcopyright{acmlicensed}

\usepackage{balance}       
\usepackage{graphics}      
\usepackage[T1]{fontenc}   
\usepackage{txfonts}
\usepackage{mathptmx}
\usepackage[pdflang={en-US},pdftex]{hyperref}
\usepackage{color}
\usepackage{booktabs}
\usepackage{textcomp}

\usepackage{enumitem}
\usepackage{lipsum}
\setlist{nolistsep}

\usepackage{microtype}        
\usepackage{ccicons}          

\usepackage{todonotes}

\def\plaintitle{Taking Informed Action on Student Activity in MOOCs}

\def\emptyauthor{}
\def\plainkeywords{MOOC; Learning Analytics; Cluster; Survey; Metrics}

\makeatletter
\def\url@leostyle{%
  \@ifundefined{selectfont}{
    \def\UrlFont{\sf}
  }{
    \def\UrlFont{\small\bf\ttfamily}
  }}
\makeatother
\def\pprw{8.5in}
\def\pprh{11in}

\definecolor{linkColor}{RGB}{6,125,233}
\hypersetup{%
  pdftitle={\plaintitle},
  pdfauthor={\emptyauthor},
  pdfkeywords={\plainkeywords},
  pdfdisplaydoctitle=true, 
  bookmarksnumbered,
  pdfstartview={FitH},
  colorlinks,
  citecolor=black,
  filecolor=black,
  linkcolor=black,
  urlcolor=linkColor,
  breaklinks=true,
  hypertexnames=false
}

\usepackage{tikz}
\newcommand\copyrighttext{%
  \footnotesize \textcopyright~Paper Authors 2017. This is the author's version of the work. It is posted here for your personal use. Not for redistribution. The definitive Version of Record was published in L@S 2017, http://dx.doi.org/10.1145/3051457.3053971. 
  }
\newcommand\authorcopyrightnotice{%
\begin{tikzpicture}[remember picture,overlay]
\node[anchor=south,yshift=10pt] at (current page.south) {\fbox{\parbox{\dimexpr\textwidth-\fboxsep-\fboxrule\relax}{\copyrighttext}}};
\end{tikzpicture}%
}

\begin{document}

\title{\plaintitle}

\numberofauthors{1}
\author{
  \alignauthor{Ralf Teusner, Kai-Adrian Rollmann, Jan Renz\\
    \affaddr{Hasso Plattner Institute, Potsdam, Germany}\\
    \email{\{firstname.lastname\}@hpi.de}\\
    }
}

\CopyrightYear{2017} 
\setcopyright{acmlicensed}
\conferenceinfo{L@S 2017,}{April 20 - 21, 2017, Cambridge, MA, USA}
\isbn{978-1-4503-4450-0/17/04}\acmPrice{\$15.00}
\doi{http://dx.doi.org/10.1145/3051457.3053971}

\maketitle

\authorcopyrightnotice

\begin{abstract}
This paper presents a novel approach to understand specific student behavior in MOOCs.
Instructors currently perceive participants only as one homogeneous group. 
In order to improve learning outcomes, they encourage students to get active in the discussion forum and remind them  of approaching deadlines.
While these actions are most likely helpful, their actual impact is often not measured.
Additionally, it is uncertain whether such generic approaches sometimes cause the opposite effect, as some participants are bothered with irrelevant information.
On the basis of fine granular events emitted by our learning platform, we derive metrics and enable teachers to employ clustering, in order to divide the vast field of participants into meaningful subgroups to be addressed individually.
\end{abstract}




\section{Introduction}
The most striking differences when comparing MOOCs with in-class courses are the mere amount of participants enrolled in MOOCs and the absence of direct personal communication. 
These differences make it difficult to gain an intuitive perception of how well a MOOC is currently running.
While a holistic view might present that the overall quiz scores are at about 80\% and the number of support tickets is on average, it would not show that there is a specific group that issued support tickets and achieved low scores due, for example due to wording problems.

Uncovering specific groups in MOOCs is difficult. Several previous works have labelled student groups \cite{wilkowski2014student, hill2013archetypes, kizilcec2013deconstructing, coffrin2014visualizing, lingras2004interval}, but there are no best practices yet on how to separate the participants.
Additionally, it is highly debatable whether there is a common separation criteria that holds for all MOOCs with respect to their strongly varying topics, requirements and settings.
Necessary steps are therefore the distinction of potential events to separate users, combine those to relevant metrics and to provide a framework that allows real-time exploration of the course status, progress and interaction.

\section{Related Work}\label{ch:related-work}
This paper contributes to the research area Learning Analytics in MOOCs.
Recent work investigates learner motivation and activity, finding and labelling characteristic groups. 

\subsection{Learner Activity}
Part of Learning Analytics research skips motivational factors and starts at learner activity, which is represented by website usage in form of clickstream events.

Whitehill et al.~\cite{whitehill2015beyond} have the goal to react automatically to student stopout.
In order to predict a good time to intervene, they include metrics such as the time since the last student activity, a measure for the regularity of the events, and the total number of different event types produced by a user.

Taylor et al.\cite{taylor2014likely} analyze event stream data from edX\footnote{\url{https://www.edx.org/}} courses and aim to predict stopout one week in advance.
Among their metrics are the total number and the average length of forum posts, the total time spent on all resources, and a correctness percentage for homework assignments.

Another approach by Halawa et al. \cite{halawa2014dropout} uses binary features to predict student dropout.
Their features include whether an assignment or a video was skipped, whether a student is lagging behind by more than two weeks, as well as whether the average quiz score fell below 50\%.

\noindent Kizilcec et al. \cite{kizilcec2013deconstructing} examine learner disengagement.
They use engagement trajectories of students, based on assignment completion and video consumption.
They argue, that for their feature choice, favoring trends of engagement over student scores was a deciding factor in finding meaningful groups.
As a second step, they test these activity groups for correlations with another set of features.
Among these features are survey results, such as enrollment intentions and overall course experience, and the number of forum posts per student.

\subsection{Characteristic Groups}\label{subsec:rel:characteristic-groups}
Previous work has derived several characteristic groups from student activity.
All of the regarded works describe their groups as mutually exclusive for a given point in time, but students may move between groups during the time of the course.
Wilkowski et al. \cite{wilkowski2014student} identify four groups of students based on their stated intentions.
Hill \cite{hill2013archetypes} describes five types of student activity.
Regarding groups with less activity, they both agree on: \emph{No-shows}, who enroll for the course, but never log in or engage with the content; \emph{Observers}, who drop in, only to see how the course is taught.
Within more active learners, Wilkowski et al. describe two groups:
\emph{Casual learners}, who engage with the content to learn a few new things related to school, work, or simply curiosity, and \emph{Completers}, who complete all necessary tasks and earn a certificate of completion.
Here, Hill sees three additional groups, \emph{Drop-ins}, who watch videos for selected topics, browse or participate in the forum, but do not attempt to complete the course;
\emph{Passive Participants}, who view the course as content to consume, participate, but do not engage with the assignments; 
\emph{Active Participants}, who take part in discussion forums and finish the majority of the assignments.

Knowing these findings, similarities can be observed in the 4 groups found by Kizilcec et al. \cite{kizilcec2013deconstructing} through clustering engagement trajectories. 
\emph{Auditing} and \emph{Completing} users seem to closely resemble the \emph{Passive Participants} and \emph{Active Participants} by Hill. The group of \emph{Sampling} users is similar to \emph{Observers}.  
 Kizilcec's \emph{Disengaging} group is not examined extensively in the descriptions of Wilkowski et al. and Hill, probably because they did not focus on changes in activity over time.

Coffrin et al. \cite{coffrin2014visualizing} define three groups on a weekly basis depending on student participation and success:
\emph{Auditors} watched videos but did not participate in assessments for a particular week;
\emph{Active} participated in assessments for a particular week;
\emph{Qualified} watched a video or participated in an assessment for a particular week and obtained marks above the 60\textsuperscript{th} percentile, leading to the assumption that these students have the capabilities to complete the course.
\emph{Auditors} and \emph{Active} are similarly defined to Kizilcec's \emph{Auditing} and \emph{Completing} groups.
In their work, Coffrin et al. also consider visualizing state changes between these groups and argue that those visualizations may benefit course instructors.

Lingras et al. \cite{lingras2004interval} analyze data from an online course offered internally to students of a particular university (non-MOOC).
They define three student groups:
\emph{Studious} download current reading material for a week as they usually study using class notes;
\emph{Crammers} download a large set of reading material, indicating their plan for a pre-test cramming;
\emph{Workers} continiously work on assignments and access the discussion forum.

Some efforts focus on single specific groups.
\hbox{Beaudoin} \cite{beaudoin2002learning} suggests that learning often happens on course absence and names this student group \emph{Invisible}, who do not show visibility on the platform in form of written contributions in the forums. In contrast, Huang et al. \cite{huang2014superposter} look at characteristics and influence of very active forum participants and label those \emph{Superposters}, who are among the top 5\% of students based on the number of forum contributions.

The various terms coined in the different works all describe related behavior and are based on similar observations and metrics.
However, there are no standardized definitions yet, thus it is helpful to compare and align the existing terms.
Clustering users should be done starting from the actual platform data instead of an artificial metric defined beforehand.
Thus we describe the verbs used on our platform and the actions they reflect in fine granularity, in order to ease portability and reproducability of the underlying concepts.

\begin{figure}[h]
  \centering
  \includegraphics[width=0.8\linewidth]{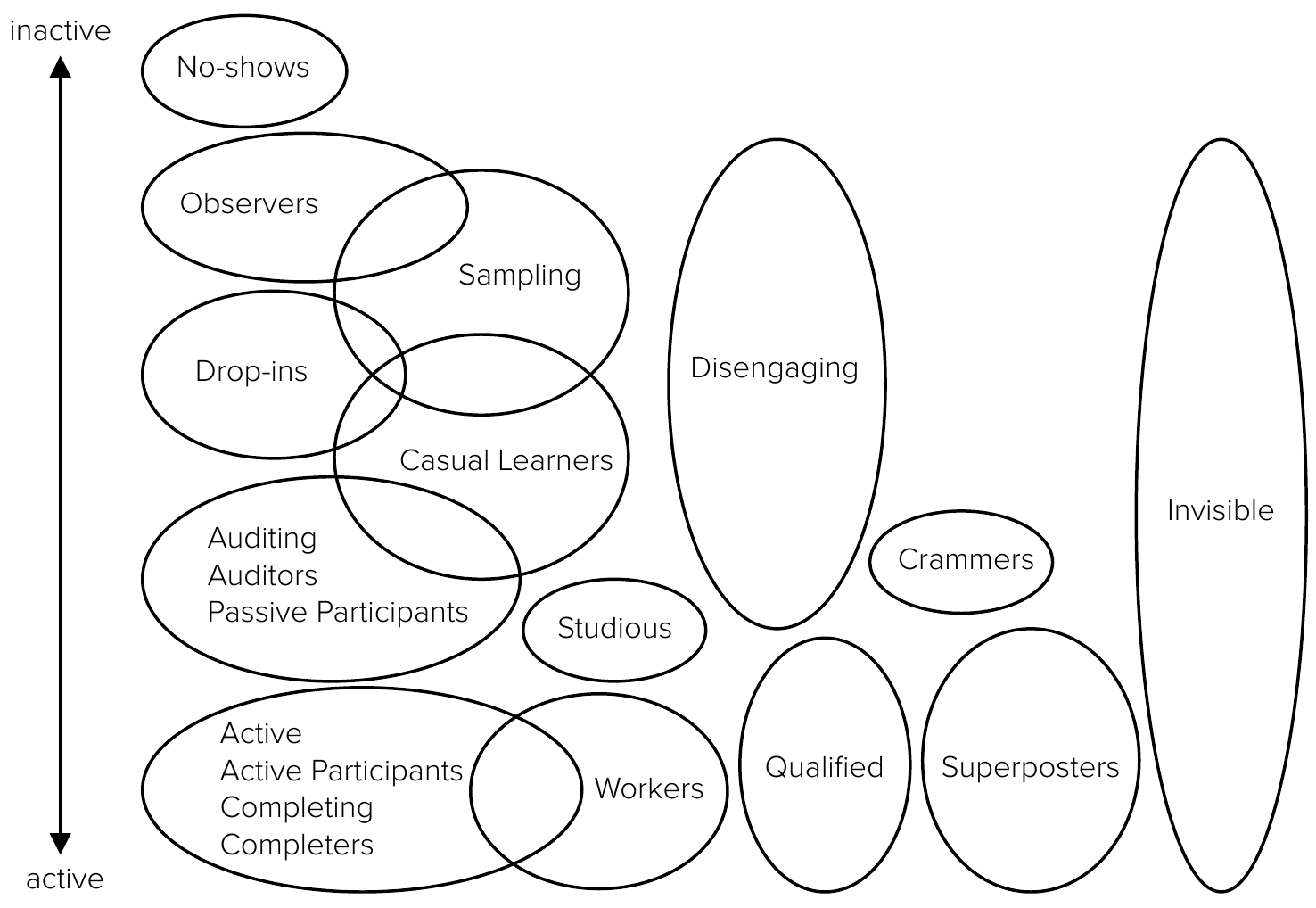}
  \caption{Approximate relations between the characteristic groups discovered by previous works.}
  \label{fig:related-groups}
\end{figure}

To illustrate the relation between the cited groups, the diagram in Figure \ref{fig:related-groups} was created.
Non-overlapping ovals indicate groups that were distinctly or differently defined, overlapping ovals signal similar definitions, and groups in the same oval were likely to be merely differently labeled.
While the width of the ovals has no further meaning, the Y-axis describes the degree of course activeness for a group.

Previous work supports course instructors and researchers in finding previously described and partly defined student groups in MOOCs.
Currently lacking is support and tooling to tailor and discover previously unknown groups.
This work presents concepts as well as an implementation to fill this current gap.


\section{Concept}\label{sec:concept}
This paper presents a concept that enables instructors of online courses to take informed action based on student activity.

On the most abstract level, our concept consists of two motivations:
(1) To find a reason for action, instructors need to be able to understand their students.
(2) After gaining a detailed understanding, instructors need to be able to take action that is as targeted and as measurable as possible.
We refer to this kind of action as  \emph{informed action}.
With the first goal in mind - understanding students - we introduce metrics that reflect student activity, some of which have also been found in previous work.
These metrics are sorted into five categories and set in relation to each other.
To approach the second goal - taking informed action - we define three categories of instructor actions and allow the instructors to \emph{encourage} students on an informed basis.

\subsection{Metrics}\label{sec:metrics}
In order to understand characteristic student activity, we collect platform usage data in the form of events triggered when users perform tracked actions.
The structure of the gathered events is similar to the definitions in the Experience API\footnote{\url{https://github.com/adlnet/xAPI-Spec/blob/master/xAPI.md}} \cite{del2013learning}.
On the gathered events, aggregations can be performed.

\begin{figure}[h]
  \includegraphics[width=\linewidth]{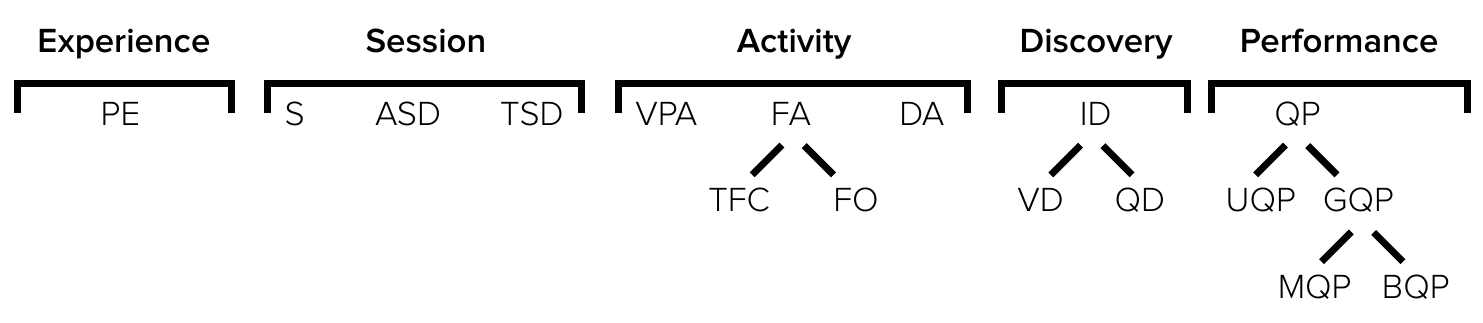}
  \caption{Overview of the metrics and their relation to higher-level platform usage categories.}
  \label{fig:metrics-overview}
\end{figure}

Yet, combining information inherent in events into possibly more meaningful metrics could provide a more abstract understanding of the underlying activity.
Therefore, we derive 17 metrics from the events and grouped them into five categories (Fig. \ref{fig:metrics-overview}). 
All metrics are computed for each individual user and specific to a particular course.
In addition to the verb counts, we allow these metrics to be used as the basis for discovering characteristic groups of students.

\begin{description}
  \item[Platform Exploration] (PE) measures the number of distinct verbs per user.
  Since most actions are possible in various courses, this metric expresses the experience a user has with the platform.
  \item[Sessions] (S) is the number of consecutive event streams (events per user have no wider gap than 30 minutes).
  \item[Total Session Duration] (TSD) is the duration of all sessions.
  \item[Average Session Duration] (ASD) is the total duration of all sessions divided by the amount of sessions.
  \item[Forum Activity] (FA) represents the sum of textual forum contribution (TFC, questions, comments, and answers) and forum observation (FO, visits and subscriptions).
  \item[Video Player Activity] (VPA) represents the sum of video player-related events (video played, paused, resized, fullscreen triggered, speed changed).
  \item[Download Activity] (DA) represents the sum of downloads.
  \item[Item Discovery] (ID) measures the share of visited items (quizzes (QD) and videos (VD)).
  \item[Quiz Performance] (QP) measures the average percentage of correct answers over all graded (GQP) and ungraded (UQP) quizzes taken. Graded quizzes are further divided into main quizzes (MQP, mandatory) and bonus quizzes (BQP, optional).
\end{description}

\subsection{Group Discovery}\label{sec:group-discovery}
Once computed, the previously described metrics can be used to create characteristic student groups.
The aim is to minimize the number of groups to be able to digest the clustering results, such as group sizes, coherencies and attributes, but to maximize a group's expressivity.
We suggest a group is very expressive when teachers can easily understand who is part of the group and are able to assign the group a label that describes their activity.
Simply assigning and labelling groups based on multiples of standard deviations from the mean for a particular metric, is not expressive, while groupings with labels like \emph{frequent video downloaders} or \emph{moderate quiz performers} would be.
The task of finding groups can be approached by classification algorithms.
While many machine learning algorithms for classification need a ground truth or other prior knowledge about expected classes (supervised learning), cluster analysis is one way to perform classification when there is little known about the data or the resulting groups (unsupervised learning).
As we do not have prior understanding about distributions or possible correlations of our metrics we decided to use clustering algorithms.

\subsection{Informed Action}\label{subsec:actions}
When course instructors have gained enough understanding about individual learner activity, they should be able to react to specific activity groups and take interventions to increase student success.
Given all options that came up, we found three action categories: \emph{Encouragement} (i.e. personalized emails), \emph{Rewarding} (i.e. badges) and \emph{Material Improvement} (i.e. add reading material, re-record videos).
\noindent From the three action categories, we focus on enabling the \emph{Encouragement} actions, since we consider them most promising to potentially influence student success.
Thus, we enable teachers to save discovered student groups and send targeted emails to members of a specific group. 

\subsection{Visualization}\label{subsec:visualization}
\begin{figure}[h]
  \includegraphics[width=\linewidth]{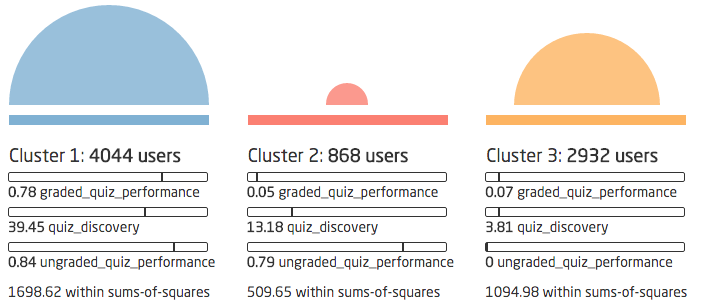}
  \caption{Cluster centers for quiz metrics measured in our MOOC.}
  \label{fig:cluster-centers}
\end{figure}

\begin{figure}[h]
  \includegraphics[width=\linewidth]{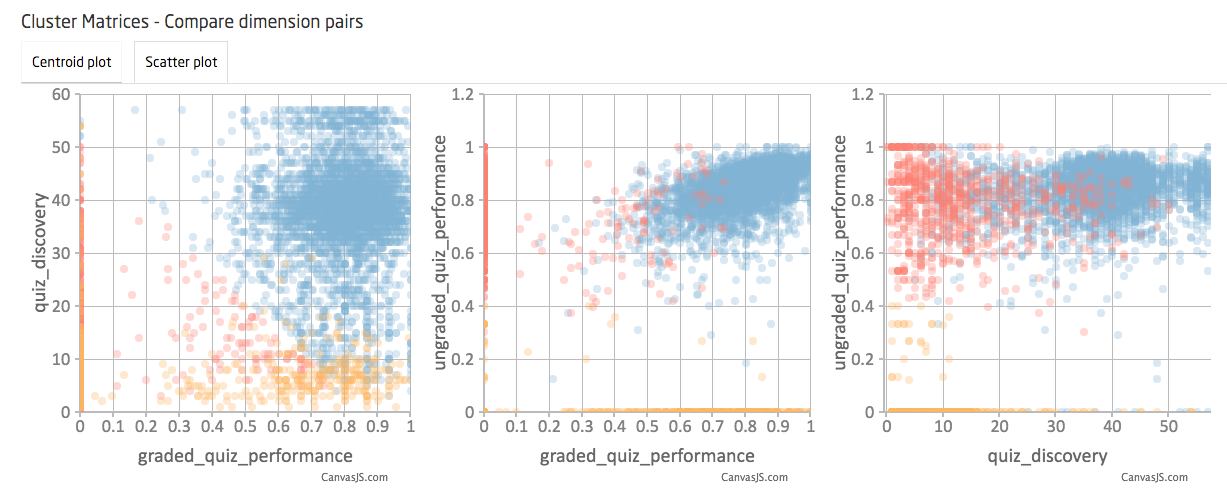}
  \caption{Scatter plot visualization.}
  \label{fig:scatter-plots}
\end{figure}

Instructors are free to choose any metrics and events the system has to offer to be included into the clustering process.
To give them a starting point, the interface also suggests several common clustering tasks to have a look into.
After the clustering has finished, several visualizations are presented, including a representation of the cluster sizes and qualities (Figure \ref{fig:cluster-centers}), the centroids, several scatter plots (Figure \ref{fig:scatter-plots}) and distribution charts.
Colors, indicating the clusters, are coherent across all visualizations, so that the graphs support each other and instructors can get a quick glimpse whether their chosen input parameters result in a meaningful distribution.
If they are confident with their findings and believe they understand the students, they can assign names to the found clusters and perform an informed action on users in the clusters afterwards.

\subsection{Performing Actions}
From the mentioned possibilities to react to student behavior, our tool Cluster Viewer allows to send targeted emails to found clusters, or half of the cluster for A/B testing.
Changes in the metrics can be compared on the respective group pages to track the effects.

\section{Evaluation}
We evaluated our concepts and tool with two interview series, ensuring we covered instructors' needs and provided a helpful tool.
The first interviews conducted were aimed to validate the acceptance of our tool and used an early stage prototype of the Cluster Viewer, allowing us to adapt the software and process if necessary.
For this series, we interviewed eight instructors from five different MOOCs, about 30 minutes each.
Regardless of knowing the cited publications concerning the different student groups, most instructors had encountered and were able to name groups as ``no-shows'', students who finish but don't show up in the forum (we coined them ``private passers'') and individuals that contribute extensive and helpful forum posts (``thoughtful thread starters'').
When offering instructors the possibility to uncover and react to students with specific behavior, they were most interested in students who: are likely to drop out (stopouts), behave malicious in the forum (trolls), are most active (actives), have questions but don't ask in the forum (reluctants), are active but don't perform well (effortlers) and those who perform best (high-performers).

A second series of interviews was conducted with five different instructors of a german course about internet security at course mid to evaluate the final implementation of the prototype.
Four of five rated the tool as helpful, while the one expressing that it did not help was confident that his existing experience was enough to support and steer a MOOC.
The four instructors were able to find interesting student groups and wanted to react to them:
Students endangered of stopout after they performed worse in graded quizzes than in ungraded quizzes should be encouraged to ask their questions in the forum prior to the next assignments.
Students performing below average and learning in few very long session should be encouraged to try another studying schedule consisting of more but shorter sessions.


\section{Future Work}\label{ch:future-work}

The methods targeting both aspects of our initial motivation - understanding students and taking informed action - can be improved individually.
The metrics could be expanded to cover aspects such as peer assessments to determine social behavior and learning styles. They could also be computed for specific weeks to reveal student trajectories.
Incorporating user optional profile data (age, gender, educational background) also adds further potential.
The exploratory data analysis can be extended to allow filtering on discovered groups and to add additional metrics to be tracked.
To further improve performance, sampling and selective rendering could provide faster feedback for the instructors.
To reproduce the findings of related work, a direct next step is to use our metrics and attempt to find previously discovered groups by others (see Subsection \ref{subsec:rel:characteristic-groups}).
As soon as experience with typical student activity has been gathered across several courses, it will be possible to highlight activity out of the norm in a running course and suggest actions for course instructors.


\section{Conclusion}
This paper presented a concept to take informed action on student activity in MOOCs.
We related our work to recent research and contributed a holistic overview of characteristics groups discovered in previous works.
Based on individual events and 17 combined metrics that may be used in any online learning platform, our prototype Cluster Viewer allows to explore student behavior within courses.
We showcased parts of our visualization and explained how instructors can send targeted emails to groups, based on their findings.
The acceptance and perceived usefulness of our tool was validated with several interviews.
A first test in a live course revealed an effect on our conducted informed action, but requires further re-evaluations in order to show statistic significance.

\balance{}

\bibliographystyle{SIGCHI-Reference-Format}
\bibliography{sample}

\end{document}